\documentclass[11pt,epsf]{article}

\usepackage{epsfig}
\usepackage{subfigure}
\usepackage{amssymb}
\usepackage{graphicx}
\usepackage{color}

\begin{document}

\baselineskip 6mm
\renewcommand{\thefootnote}{\fnsymbol{footnote}}


\newcommand{\nc}{\newcommand}
\newcommand{\rnc}{\renewcommand}



\newcommand{\tcb}{\textcolor{blue}}
\newcommand{\tcr}{\textcolor{red}}
\newcommand{\tcg}{\textcolor{green}}


\def\be{\begin{eqnarray}}
\def\ee{\end{eqnarray}}
\def\nn{\nonumber\\}


\def\ct{\cite}
\def\la{\label}
\def\eq#1{(\ref{#1})}


\def\a{\alpha}
\def\b{\beta}
\def\g{\gamma}
\def\G{\Gamma}
\def\d{\delta}
\def\D{\Delta}
\def\e{\epsilon}
\def\et{\eta}
\def\ph{\phi}
\def\Ph{\Phi}
\def\ps{\psi}
\def\Ps{\Psi}
\def\k{\kappa}
\def\l{\lambda}
\def\L{\Lambda}
\def\m{\mu}
\def\n{\nu}
\def\th{\theta}
\def\Th{\Theta}
\def\r{\rho}
\def\s{\sigma}
\def\S{\Sigma}
\def\ta{\tau}
\def\o{\omega}
\def\O{\Omega}
\def\pr{\prime}


\def\half{\frac{1}{2}}

\def\goto{\rightarrow}

\def\na{\nabla}
\def\grad{\nabla}
\def\curl{\nabla\times}
\def\div{\nabla\cdot}
\def\pa{\partial}
\def\fr{\frac}

\def\bra{\left\langle}
\def\ket{\right\rangle}
\def\lb{\left[}
\def\lc{\left\{}
\def\ls{\left(}
\def\lp{\left.}
\def\rp{\right.}
\def\rb{\right]}
\def\rc{\right\}}
\def\rs{\right)}

\def\vac#1{\mid #1 \rangle}


\def\td#1{\tilde{#1}}
\def\check{ \maltese {\bf Check!}}


\def\Tr{{\rm Tr}\,}
\def\det{{\rm det}}
\def\text#1{{\rm #1}}


\def\bc#1{\nnindent {\bf $\bullet$ #1} \\ }
\def\ch {$<Check!>$ }
\def\ss {\vspace{1.5cm}}
\def\inf{\infty}

\begin{titlepage}

\hfill\parbox{5cm} { }

\vspace{25mm}

\begin{center}
{\Large \bf RG flow of entanglement entropy to thermal entropy}

\vskip 1. cm
   {Ki-Seok Kim$^{a}$\footnote{e-mail : tkfkd@postech.ac.kr }
   and Chanyong Park$^{a,b}$\footnote{e-mail : chanyong.park@apctp.org}}

\vskip 0.5cm

{\it $^a\,$ Department of Physics, Postech, Pohang, Gyeongbuk 790-784, Korea }\\
{\it $^b\,$ Asia Pacific Center for Theoretical Physics, Pohang, Gyeongbuk 790-784, Korea } \\

\end{center}

\thispagestyle{empty}

\vskip2cm


\centerline{\bf ABSTRACT} \vskip 4mm

Utilizing the holographic technique, we investigate how the entanglement entropy evolves along the RG flow. After introducing a new generalized temperature which satisfies the thermodynamics-like law even in the IR regime, we find that the renormalized entropy and the generalized temperature in the IR limit approach the thermal entropy and thermodynamic temperature of a real thermal system. This result implies that the microscopic quantum entanglement entropy in the IR region leads to the thermodynamic relation up to small quantum corrections caused by the quantum entanglement near the entangling surface. Intriguingly, this IR feature of the entanglement entropy universally happens regardless of the detail of the dual field theory and the shape of the entangling surface. We check this IR universality with a most general geometry called the hyperscaling violation geometry which is dual to a relativistic non-conformal field theory.

\vspace{1cm}

\vspace{2cm}


\end{titlepage}

\renewcommand{\thefootnote}{\arabic{footnote}}
\setcounter{footnote}{0}



\section{Introduction}

Recently, considerable attention has been paid to the entanglement entropy for understanding quantum aspects of theoretical and experimental physics. In general, a quantum system governed by a microscopic theory does not prefer any specific direction in time. So it is generally in a reversible process. For a macroscopic system, however, the irreversibility naturally occurs as the second law of thermodynamics. In the quantum information theory, it has been shown that quantum information can be thermalized via the unitary time evolution, and that there exists the link between the quantum information and the thermal entropy \cite{Popescu:2006,Sagawa:2016}. In spite of these studies, it is still unclear what kind of the underlying structure for a microscopic theory leads to such macroscopic irreversibility. In order to understand the occurrence of thermodynamics from a microscopic quantum theory, we need to study further the connection  between  the microscopic theory and thermodynamics.

In the quantum field theory, it has been known that there exists a microscopic irreversible process along the renormalization group (RG) flow, which is called the $c$-theorem. The $c$-theorem claims that the $c$-function decreases monotonically along the RG flow. It has first been proved for a two-dimensional quantum field theory and later for a four-dimensional case \cite{Zamolodchikov:1986gt,Komargodski:2011vj,Jafferis:2011zi,Myers:2010tj,Casini:2004bw,Casini:2012ei}.  In order to represent the microscopic irreversibility of a quantum system, an important physical quantity is the entanglement entropy which measures the entanglement between quantum states \cite{Solodukhin:2008dh}-\cite{Park:2015dia}. In general, it is not easy work to calculate the entanglement entropy of an interacting quantum field theory. Based on the AdS/CFT correspondence \cite{Maldacena:1997re,Gubser:1998bc,Witten:1998qj,Witten:1998zw}, recently, a new and relatively simple way to evaluate the entanglement entropy was proposed \cite{Ryu:2006bv,Ryu:2006ef}.
In the UV limit, it has been well known that the entanglement entropy allows a thermodynamics-like law \cite{Casini:2008cr,Blanco:2013joa}.
The origin of the thermodynamics-like law can be readily understood from the non-negativity of the renormalized (or relative) entropy describing the distance between two quantum states. Introducing an entanglement temperature, the thermodynamics-like law in the UV region can be rewritten as \cite{Bhattacharya:2012mi}-\cite{Ben-Ami:2015zsa}
\be			\la{res:thermolaw}
\D {S} \approx \fr{\D E}{T_E} + \cdots.
\ee
where the ellipsis means higher order corrections relying on the small subsystem size. Ignoring all higher order corrections, the entanglement temperature, $T_E$, shows a universal behavior inversely proportional to the subsystem size. Despite the similarity to the real thermodynamic law, the entanglement temperature is not a genuine thermodynamic quantity because the real thermodynamic temperature is independent of the system size. Thus, it would be interesting to ask what the meaning of the entanglement temperature is and how it can be connected to the real temperature. In this work, we will investigate how the macroscopic thermodynamics law can emerge from the microscopic quantum system along the RG flow.

Recently, it has been shown that the above thermodynamics-like law plays a central role in reconstructing the linearized Einstein equation of the dual geometry only from CFT data \cite{Swingle:2014uza,Lashkari:2013koa,Faulkner:2013ica,Nozaki:2013vta}. However, the thermodynamics-like law and entanglement temperature defined above are valid only in the UV region because of neglecting all higher order corrections. In order to go beyond the linearized level and to investigate the RG flow correctly, it is required to define new quantities which are valid even in the IR regime. By involving all higher order corrections, we can define a generalized thermodynamics-like law such that it is valid in the entire RG scale. Even in the UV region the higher order corrections usually modify the entanglement temperature, so that we all need to generalize the entanglement temperature which we simply call a generalized temperature from now on. Note that these generalized concepts are inevitable for representing the RG flow of the entanglement entropy correctly.

Defining a renormalized entropy by subtracting the ground state entanglement entropy, it does not suffer from the UV divergence any more and its  generalized thermodynamics-like law reproduces the previous result in \eq{res:thermolaw} in the UV region. In the IR limit, intriguingly, we find that the renormalized entropy  has a universal form regardless of the microscopic detail and the shape of the entangling surface
\be
\bar{S} = S_{th} + S_q ,
\ee
where $S_{th}$ and $S_q$ indicate a thermal entropy and a small quantum correction, respectively. In this case, the thermal entropy is nothing but the Bekenstein-Hawking entropy of the dual black hole geometry. Note that the similar structure also appears in black hole physics \cite{Cardy:1986ie}-\cite{Susskind:1993ws}. The above universal form of the IR entanglement entropy implies that the excited state entanglement entropy can be thermalized in the IR limit with the small quantum correlation near the entangling surface. Intriguingly, we find that the generalized temperature also approaches the thermodynamic temperature in the IR limit. This fact implies that the generalized thermodynamics-like law we defined leads to the real thermodynamic law in the IR limit up to the small quantum correction. In this work, we holographically check the universal feature of the IR entanglement entropy in conformal and non-conformal field theories with a strip- or ball-shaped entangling surface.

The rest of this paper is organized as follows: In Sec. 2, we first look into the RG flow of the entanglement entropy for a two-dimensional CFT. In this case, since we can find an analytic form of the entanglement entropy in the entire RG scale, we explicitly shows how the renormalized entropy and generalized temperature approach the thermal entropy and the real thermodynamic temperature along the RG flow. In Sec. 3, we take into account a higher dimensional CFT in which the entangling surface can have various different shapes. We find that the thermal entropy gives the main contribution to the IR entanglement entropy regardless of the dimension and topology of the entangling surface. We further show in Sec. 4 that this IR feature of the entanglement entropy universally appears even in the non-conformal field theory. We finish this work with some concluding remarks in Sec. 5.


\section{RG flow of the entanglement entropy}

Although the entanglement entropy is well defined in a general quantum field theory, it is not easy work to evaluate the entanglement entropy of an interacting quantum field theory. However, the recent holographic studies proposed a new way to calculate the entanglement entropy on the dual gravity side \cite{Ryu:2006bv,Ryu:2006ef}. Based on the AdS/CFT correspondence \cite{Maldacena:1997re,Gubser:1998bc,Witten:1998qj,Witten:1998zw}, the holographic entanglement entropy can be obtained by evaluate the area of the minimal surface extended to the corresponding dual geometry. Through this holographic technique, in this work, we try to understand how the entanglement entropy evolves along the RG flow. In general, since the entanglement entropy crucially relies on the dimension and shape of the entangling surface dividing a total system into two subsystems, we first concentrate on a three-dimensional AdS space which is dual to a two-dimensional CFT. In this case, the entangling surface is just a point, so the subsystem is always given by an interval on a line. The black hole metric in a three-dimensional AdS space is represented as
\be			\la{res:Lifshitzmetric}
ds^2 = - \fr{R^{2  }}{z^{2 }} f(z) dt^2 + \fr{R^2}{z^2 f(z)} dz^2 + \fr{R^2}{z^2} dx^2 ,
\ee
with the following black hole factor
\be
f(z) = 1 - \fr{z^{2}}{z_h^{2}} .
\ee
Following the AdS/CFT correspondence, this black hole geometry can be matched to the two-dimensional CFT at finite temperature. More precisely, black hole quantities can be identified with macroscopic quantities classifying thermodynamics of the dual field theory. Setting $R=1$ for simplicity, the Hawking temperature and Bekenstein-Hawking entropy contained in the boundary volume, $-l/2 \le x \le l/2$, are given by
\be
T_H &=& \fr{1}{2 \pi z_h}   , \la{res:blackholeentemp} \\
S_{th} &=& \fr{1}{4 G} \fr{l}{z_h} ,
\ee
and the first law of thermodynamics, $dE = T_H d S_{th}$, leads to the internal energy
\be			\la{res:Lifshitzenergy}
E = \int T_H d S_{th} =\fr{1}{16 \pi  G} \fr{l}{z_h^{2}} = \fr{1}{2} T_H S_{th} .
\ee
Note that the Hawking temperature is global in that it does not depend on the system size, $l$.

Now, let us study the entanglement entropy of the corresponding CFT. If we take a subsystem located at $-l/2 \le x \le l/2$. the holographic entanglement entropy is determined from
\be			\la{act:minimalsurface}
S_E = \fr{1}{4G} \int_0^{l/2} dx  \ \fr{R}{z} \sqrt{1 + \fr{z'^2}{f}} ,
\ee
where the prime means a derivative with respect to $x$. Due to the absence of the  explicit $x$-dependence, there exists a well-defined conserved quantity
\be
H = - \fr{R}{z} \fr{1}{\sqrt{1 + z'^2 /f}} .
\ee
In addition, the $Z_2$ symmetry ($x \to - x$) allows a turning point at $x=0$ and, at the same time, the smoothness of the minimal surface at the turning point leads to $z'=0$. These constraints determine the subsystem size and the holographic entanglement entropy in terms of the turning point denoted as $z_0$
\be			\la{eq:subsystemsize}
l &=& 2 \int_0^{z_0} dz \ \fr{z}{\sqrt{f} \sqrt{z_0^2 - z^2}} , \\
		\la{eq:entanglementen}
S_E &\equiv& \fr{A}{4 G} = \fr{1}{2 G} \int_\e^{z_0} dz \fr{z_0}{z \sqrt{f} \sqrt{z_0^2 - z^2}} ,
\ee
where a UV cutoff $\e$ corresponding to the lattice spacing is introduced for the regularization. When $f=1$ with $z_h = \infty$, the above black hole geometry reduces to a pure AdS space corresponding to the ground state of the dual CFT. In this case, the turning point is located at $z_0= l/2$ and the ground state entanglement entropy reads
\be      \la{res:groundent}
S_g=  \fr{1}{2 G} \log \fr{l}{\e} .
\ee
Noting that the Newton constant is associated with the central charge of the dual CFT, $c = \fr{3 R}{2 G}$, it is perfectly matched to that obtained from a two-dimensional CFT \cite{Calabrese:2004eu,Calabrese:2005zw,Calabrese:2009qy}.

For $f\ne1$, the black hole geometry can be well described by the previous thermodynamic quantities, which may be understood as IR physics caused by thermalization of the quantum excitations. In the UV region, the holographic entanglement entropy derived in the black hole geometry can be regarded as the excited state entanglement entropy and, from the RG flow point of view, this microscopic quantity must continuously connected to the macroscopic one. Performing the integrals in \eq{eq:subsystemsize} and \eq{eq:entanglementen} and rewriting the entanglement entropy in terms of the subsystem size, we can find the analytic form of the entanglement entropy
\be
S_E  = \fr{1}{2 G}  \log \ls \fr{\b}{\pi \e}  \sinh \ls \fr{\pi l}{\b}\rs \rs ,
\ee
where $\b = 1/ T_H$. This is the exact same as the entanglement entropy obtained in a two-dimensional CFT at finite temperature \cite{Calabrese:2004eu,Calabrese:2005zw,Calabrese:2009qy}. Following the previous RG flow prescription, this excited state entanglement entropy should have a connection to the thermal entropy in the IR region. From now on, we discuss about how the thermal entropy can occur from the entanglement entropy along the RG flow.

In order to see the connection between the entanglement and thermal entropies, we should first define a renormalized (or relative) entropy by subtracting the ground state entanglement entropy
\be
\bar{S} \equiv S_E - S_g  = \fr{1}{2 G}  \log \ls \fr{2 z_h}{l}  \sinh \ls \fr{l}{2 z_h}\rs \rs ,
\ee
which is required to remove unphysical UV divergences. In the UV region ($l \ll z_h$), it has been known that the renormalized entropy satisfies the thermodynamics-like law after introducing an appropriate entanglement temperature, which is universally proportional to the inverse of the subsystem size \cite{Bhattacharya:2012mi}. This fact becomes clear when expanding the above renormalized entropy in the UV region
\be
\bar{S}  &=& \fr{1}{48 G}  \fr{l^2}{z_h^2}   + {\cal O} \ls l^4\rs  .
\ee
Ignoring higher order corrections and applying the thermodynamics-like law, $E \approx T_E \bar{S} /2$, the entanglement temperature reads \cite{Bhattacharya:2012mi,Park:2015hcz,Park:2015afa}
\be		\la{res:penttemp}
T_E = \fr{6} {\pi l} .
\ee
This result shows the universal behavior mentioned before and describes how the excitation energy increases the entanglement entropy. The entanglement temperature defined in the UV region is totally different from the thermodynamic temperature because the thermodynamic temperature is independent of the system size. Therefore, the thermodynamics-like law of the entanglement entropy has nothing to do with the real thermodynamic law. However, this UV story dramatically changes in the IR regime where the real thermodynamic law can universally emerge from the IR entanglement entropy regardless of the microscopic details and the shape of the entangling surface.

To see the universal IR feature for the above two-dimensional CFT, it should first be noted that when deriving \eq{res:penttemp}, all higher order corrections of the renormalized entropy were ignored. This fact implies that the entanglement temperature defined in \eq{res:penttemp} is valid only in the UV region. In order to represent the RG flow correctly, we need to generalize the entanglement temperature to be applicable even in the IR regime. We define a generalized temperature as
\be			\la{def:entanglementtemp}
\fr{1}{\bar{T}} \equiv \half \fr{\bar{S}}{ \bar{E}} = \fr{4 \pi  z_h^2}{l}\log \ls \fr{2 z_h}{l}  \sinh \ls \fr{l}{2 z_h}\rs \rs ,
\ee
where all higher order corrections of the renormalized entropy are involved. Then, the generalized thermodynamics-like law, $E =  {\bar{T}} \bar{S}/2$, becomes an exact relation satisfied in the entire region of $l$, including the IR $(l \to \infty)$ as well as UV $(l \to 0)$ region. In the UV region, the renormalized entropy and the newly defined generalized temperature are expanded into
\be
\bar{S}  &=& \fr{1}{48 G}  \fr{l^2}{z_h^2}  \ls 1 - \fr{l^2}{120 z_h^2} + \cdots   \rs  , \\
 \fr{1}{\bar{T}} &=& \fr{1}{T_E} \ls 1 - \fr{l^2}{120 z_h^2} + \cdots   \rs ,
\ee
where $T_E$ appears at leading order, as expected. Recently, there was an interesting study to reconstruct the linearized dual geometry from the thermodynamics-like law represented by $T_E$ \cite{Swingle:2014uza,Lashkari:2013koa,Faulkner:2013ica,Nozaki:2013vta}. In order to go beyond the linearized level, the generalized temperature defined here may play an important role, because it contains all higher order information related to
the inner region of the dual geometry. Therefore, it would be interesting to reconstruct a nonlinear Einstein equation from the generalized thermodynamics-like law. We leave it as a future work.

To understand the connection between the entanglement and thermal entropies, we look into the renormalized entropy in the IR limit ($l \to \infty$) where $z_0$ approaches $z_h$. In the IR region, the renormalized entropy has the following expansion
\be
\bar{S}  = \fr{1}{4 G} \fr{l}{z_h} - \fr{1}{2 G} \log \ls \fr{l}{4 z_h} \rs +  {\cal O} \ls \e^{- l/z_h} \rs ,
\ee
where the leading contribution is the exact same as the thermal entropy. Intriguingly, this result indicates that the quantum entanglement entropy in the UV region, which has no any connection to the thermal entropy, can evolve into the macroscopic thermal entropy along the RG flow. For more understanding, let us consider the inverse of the generalized temperature, $\bar{\b}  = 1 /\bar{T}$, which represents manifestly how the entanglement entropy is thermalized in the IR regime. In the UV regime, the leading behavior of $\bar{\b} $ is linearly proportional to $l$, so that the entanglement entropy describes quantum aspects rather than thermal ones. In the IR regime, on the other hand, $\bar{\b} $ behaves as
\be
\bar{\b}  = 2 \pi z_h + \fr{4 \pi z_h^2}{l} \log \ls \fr{z_h}{l} \rs  + \cdots ,
\ee
and approaches the inverse of the Hawking temperature. In Fig. 1(a), we depict how the entanglement temperature continuously approaches the Hawking temperature in the IR regime. In order to characterize the thermal and quantum aspects, let us consider the RG flow of $\bar{\b} $
\be
l \fr{d \bar{\b} }{d l} = 2 \pi z_h \lb  \coth \fr{l}{2 z_h} - \fr{2 z_h}{l}  \lc 1 + \log \ls \fr{2 z_h}{l} \sinh \fr{l}{2 z_h} \rs  \rc \rb  .
\ee
Recalling that the thermodynamic temperature is independent of the system size, the RG flow of $\bar{\b} $ in Fig. 1(b) confirms that the generalized temperature really becomes the thermodynamic temperature in the IR regime. Intriguingly, Fig. 1(b) has a maximum near $l =7.019$. If the subsystem size is smaller than this critical value the system behaves like a quantum system, whereas above the critical value the system approaches a thermal system.

\begin{figure}
\begin{center}
\vspace{0cm}
\hspace{-0.5cm}
\subfigure[]{ \includegraphics[angle=0,width=0.45\textwidth]{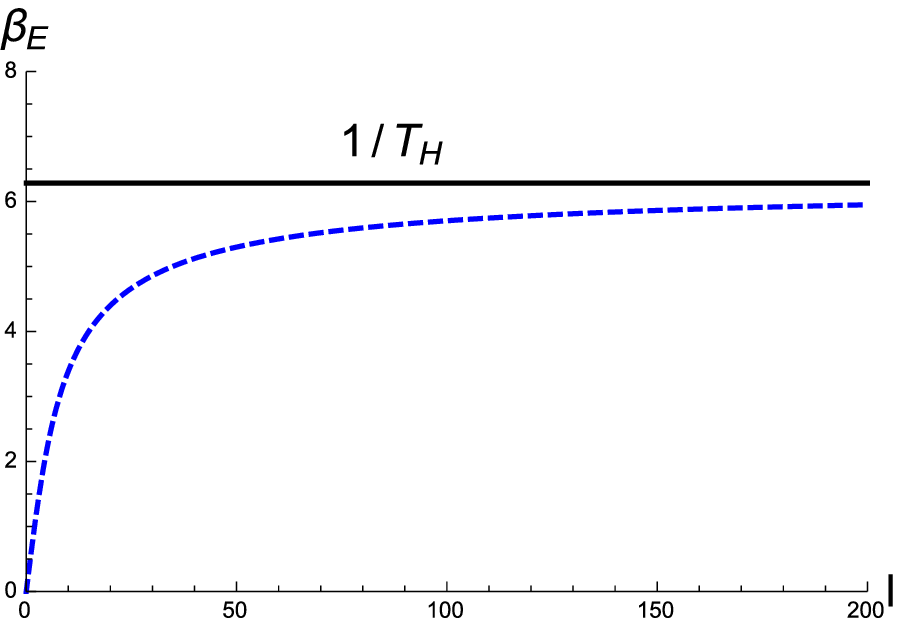}}
\hspace{0cm}
\subfigure[]{ \includegraphics[angle=0,width=0.45\textwidth]{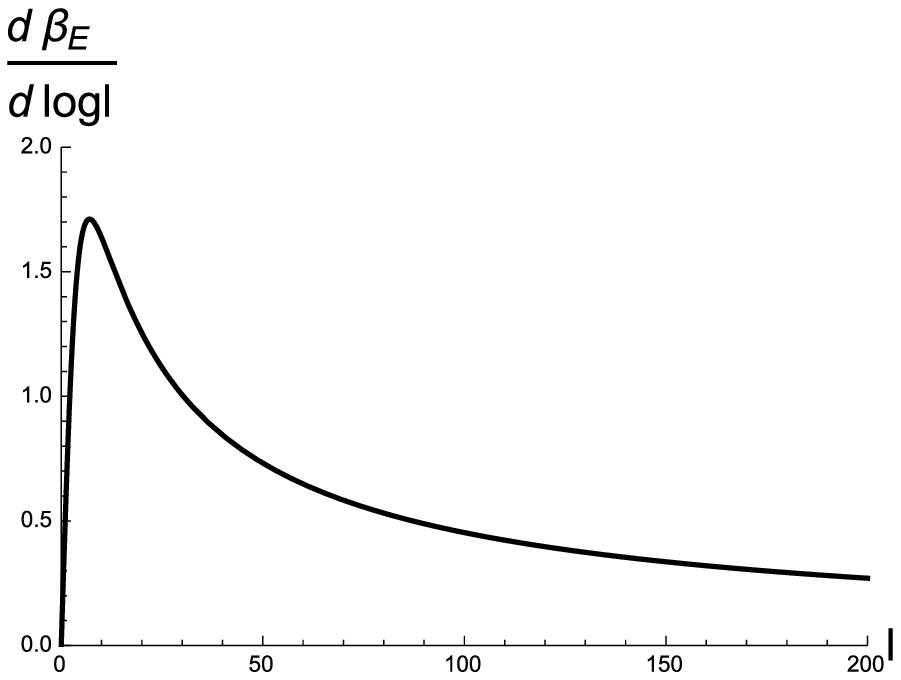}}
\vspace{-0cm}
\caption{\small (a) The inverse of the entanglement temperature depending on the subsystem size and (b) its RG flow. }
\label{number}
\end{center}
\end{figure}

For a two-dimensional CFT, in summary, we showed that the excited state entanglement entropy can be thermalized and results in the thermal entropy in the IR regime. This fact implies that the IR renormalized entropy can have the following expansion form
\be
\bar{S} = S_{th} + S_{q} ,
\ee
where $S_{q}$ indicates the quantum correction caused by the remaining short-distance quantum correlation  near the entangling surface. The emergence of the macroscopic thermal entropy from the quantum entanglement entropy requires that $S_{th}$ should be larger than $S_{q}$ at least in the IR regime. For the above two-dimensional example, $S_{th}$ increases linearly as $l$ increases, while the quantum correction increases logarithmically. The similar feature also occurs  in a higher dimensional theory regardless of the microscopic details and the topology of the entangling surface, as will be shown. This fact implies that the emergence of the thermodynamic properties in the IR region is a universal behavior of the quantum entanglement entropy governing the microscopic theory. We will check this universality in the following sections.

Before going to our discussion on the entanglement entropy above two dimensions, we would like to point out the existence of a crossover scale given by the maximum of $\frac{d \beta_{E}}{d \ln l}$ in Fig. 1(b). This crossover scale should be regarded to be a special feature detected by the entanglement entropy near quantum criticality since such an energy scale is not shown in any conventional thermodynamic properties. Certainly, it shows the change of an entanglement pattern in the ground-state wave-function near quantum criticality. We suspect that this energy scale may be involved with a full development of thermalization: Below this temperature scale, strong inelastic scattering gives rise to thermalization, where dynamic properties would be described by emergent hydrodynamic equations. Unfortunately, it is not clear at all how to verify the existence of such an energy scale in correlation functions.

\section{Quantum correction of the IR entanglement entropy}

For a higher dimensional theory unlike the two-dimensional case, the entangling surface can have various topologies and the entanglement entropy crucially depends on the shape of the entangling surface. In order to study the universality of the IR entanglement entropy, therefore, we should investigate what kind of the quantum correction appears relying on the shape of the entangling surface. In this section, we will consider a strip- and ball-shaped entangling region and  show that the IR entanglement entropy reduces to the thermal entropy regardless of the shape of the entangling surface.

\subsection{Strip-shaped entangling region}

For a $(d+1)$-dimensional AdS black hole dual to an excited state of a $d$-dimensional CFT, a general metric is given by
\be
ds^2 = \fr{1}{z^2} \ls - f(z) dt^2  + d x_i^2 + \fr{1}{f(z)}dz ^2 \rs ,
\ee
where $i$ runs from $1$ to $d-1$ and the black hole factor is given by
\be
f(z) = 1 - \fr{z^d}{z_h^d} ,
\ee
Let us first consider a strip-shaped subsystem with the following parameterization
\be
- \fr{l}{2} \le  x \le \fr{l}{2} \quad {\rm and} \quad - \fr{L}{2} \le  x_2, \cdots , x_{d-1} \le \fr{L}{2} ,
\ee
where we replaced $x_1$ by $x$ for convenience. Above $l$ and $L$ denote the size of the subsystem and total system, respectively. Then, the holographic entanglement entropy is governed by
\be
S_E = \fr{L^{d-2}}{4 G} \int_{-/2}^{l/2} dx \ \fr{\sqrt{f+z'^2}}{z^{d-1} \sqrt{f}} ,
\ee
where the prime means a derivative with respect to $x$. Denoting the turning point as $z_0$, the subsystem size and the entanglement entropy can be determined as functions of the turning point
\be
l &=& 2 \int_0^{z_0} dz \ \fr{z^{d-1}}{\sqrt{f} \sqrt{z_0^{2(d-1)} - z^{2(d-1)}}} , \la{int:stripsize}\\
S_E &=& \fr{L^{d-2}}{2 G} \int_0^{z_0} dz \ \fr{z_0^{d-1}}{z^{d-1} \sqrt{f} \sqrt{ z_0^{2(d-1)} - z^{2(d-1)}}} .
\la{int:striphee}
\ee
From these relations, the entanglement entropy in the UV region ($z_0/z_h \ll 1$) has been well studied. Similar to the previous two-dimensional case, the thermodynamics-like law in the UV region leads to the entanglement temperature proportional to the inverse of the subsystem size.

In order to define the renormalized entropy for an excited state, let us first consider the ground state entanglement entropy. To do so, we consider a pure AdS space by taking $f=1$. When the subsystem size is given by $l$, we need to introduce a new turning point because the position of the turning point relies on the background geometry. Denoting the turning point of the pure AdS geometry as $z_*$, it is determined to be
\be
l = 2 \int_0^{z_*} dz \ \fr{z^{d-1}}{\sqrt{z_*^{2(d-1)} - z^{2(d-1)}}}
= \frac{2 \sqrt{\pi } z_* \Gamma \left(\frac{d}{2 (d-1)}\right)}{\Gamma \left(\frac{1}{2  (d-1)}\right)} .
\ee
Moreover, the exact ground state entanglement entropy reads
\be                \la{res:groundstHEE}
S_g &=& \fr{L^{d-2}}{2 G} \int_0^{z_*} dz \ \fr{z_*^{d-1}}{z^{d-1} \sqrt{ z_*^{2(d-1)} - z^{2(d-1)}}} \nn
&=&\frac{ 1}{2 (d-2) G} \fr{L^{d-2}}{\epsilon ^{d-2}}
-\frac{2^{d-3} \pi ^{\frac{d-1}{2}}  \Gamma \left(\frac{d}{2 (d-1)}\right)^{d-1}}{(d-2) G \ \Gamma \left(\frac{1}{2 (d-1)}\right)^{d-1} } \fr{L^{d-2}}{ l^{d-2}}  .
\ee
Then, the renormalized entropy can be determined as
\be
\bar{S} \equiv S_E -S_g ,
\ee
which represents the difference between the ground and excited state entanglement entropies. One important thing we should note is that the renormalized entropy has no UV divergence and is independent of the renormalization scheme.

For the higher dimensional case unlike the previous two-dimensional case, since the integrals in \eq{int:stripsize} and \eq{int:striphee} do not allow the exact calculation, we focus on the IR behavior near $z_0 \approx z_h$. The integrand of \eq{int:stripsize} is singular at $z = z_0$,  so the main contribution comes from the integration near $z \approx z_0$. Using this fact, the subsystem size can be approximately written as
\be     \la{res:IRsubsysize}
l  \approx  - \fr{\sqrt{2}  z_0 }{\sqrt{d (d-1)} } \log  \ls 1  - \fr{z_0}{z_h} \rs + \cdots   ,
\ee
where the ellipsis denotes finite higher order correction. This result shows that the subsystem size diverges logarithmically when the turning point approaches the horizon.
From \eq{int:striphee}, one can see that the excited state entanglement entropy  have two kids of divergences. The first one is the UV divergence appearing at $z=0$. Since the ground state entanglement entropy cancels this divergence, this UV divergence does not appear in IR physics described by the renormalized entropy. The second divergence appears at $z=z_0$ only at $z_0 = z_h$. This is crucial for understanding IR physics because the ground state entanglement entropy has no such an IR divergence. In the IR region ($z_0 \approx z_h$), the leading contribution of \eq{int:striphee} also gives rise to a logarithmic behavior similar to \eq{res:IRsubsysize}
\be
S_E \sim  - \fr{\sqrt{2}   }{\sqrt{d (d-1)} }  \fr{L^{d-2}}{2 G} \fr{1}{z_0^{d-2}}  \log  \ls 1  - \fr{z_0}{z_h} \rs .
\ee
Rewriting it in terms of $l$, the leading contribution of the IR entanglement entropy is given by a term linearly proportional to the subsystem size.
\be
S_E \sim \fr{L^{d-2} l}{4 G z_h^{d-1}} ,
\ee
which is exactly the same as the thermal entropy of the dual field theory. This result shows, as mentioned before, that the leading behavior of the IR entanglement entropy defined in the strip-shaped region reduces to the thermal entropy.

In order to check the universal behavior of the IR entanglement entropy more precisely, we need to confirm whether the remaining quantum correction of the IR entanglement entropy is really smaller than the thermal entropy. To do so, it is worth noting that one can rewrite the renormalized entropy as the following form
\be        \la{res:stripenten}
\bar{S} &=& \fr{L^{d-2} }{2 G }  \int_0^{z_0} dz \ \fr{z^{d-1}}{\sqrt{f} \sqrt{z_0^{2(d-1)} - z^{2(d-1)}}} + \fr{L^{d-2} }{2 G } \int_0^{z_0} dz \ \fr{\sqrt{ z_0^{2(d-1)} - z^{2(d-1)}}}{z_0^{d-1} z^{d-1} \sqrt{f} } - S_g .
\ee
When $z_0 \to z_h$ the first integral gives rise to the leading thermal entropy discussed above, while the remaining terms correspond to the quantum correction. One can easily check that the remaining quantum correction is finite in the IR limit ($z_0 = z_h$). Due to the finiteness of this quantum correction in the IR region, the renormalized entropy can be perturbatively expanded into
\be      \la{res:REEstrip}
\bar{S} = \fr{L^{d-2} l}{4 G z_h^{d-1}}  + a_0 + a_{UV} \ l^{- (d-2)} +  a_1 \ l^{-b_1} + a_2 \ l^{-b_2} \cdots,
\ee
where $0< b_1 < b_2 < \cdots$ and $a_{UV}$ is fixed from the ground state entanglement entropy
\be
a_{UV} = \frac{2^{d-3} \pi ^{\frac{d-1}{2}}  \Gamma \left(\frac{d}{2 (d-1)}\right)^{d-1}}{(d-2) G \ \Gamma \left(\frac{1}{2 (d-1)}\right)^{d-1} } L^{d-2} .
\ee

\begin{figure}
\begin{center}
\vspace{0cm}
\hspace{-0.5cm}
\subfigure[]{ \includegraphics[angle=0,width=0.4\textwidth]{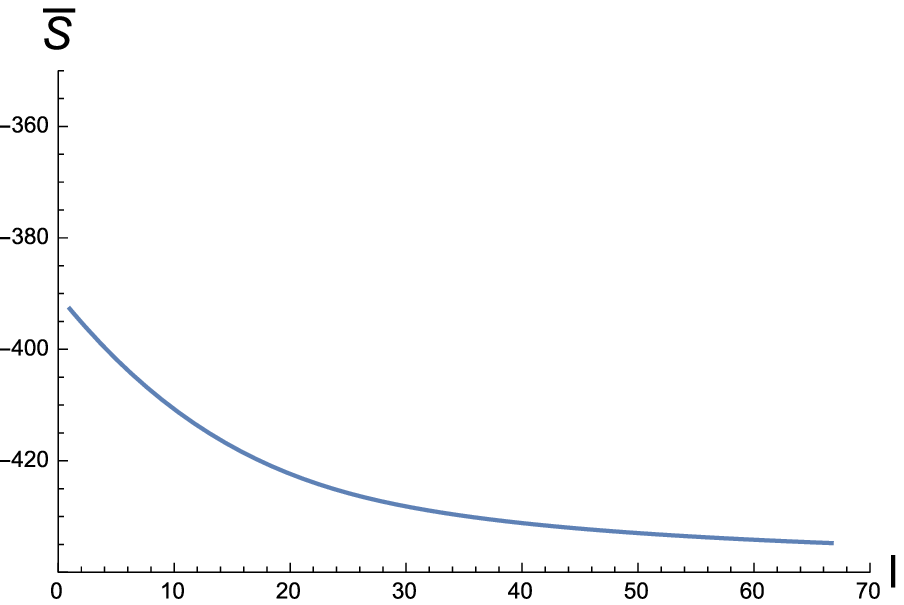}}
\hspace{0.5cm}
\subfigure[]{ \includegraphics[angle=0,width=0.4\textwidth]{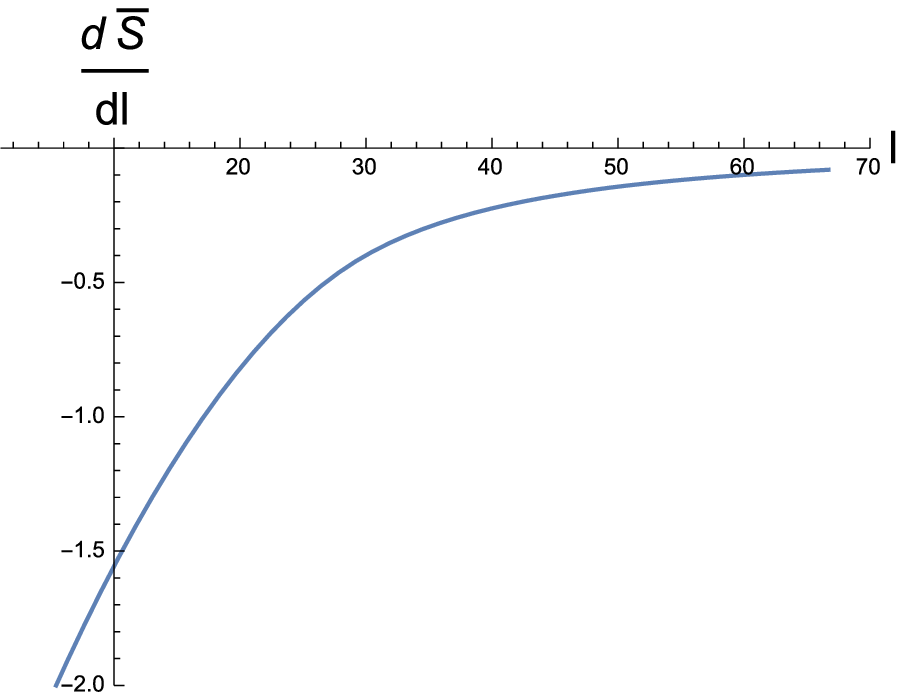}}
\vspace{-0cm}
\caption{\small The renormalized entropy and  its derivative depending on the subsystem size. Fig 1(a) shows that the renormalized entropy monotonically decreases along the RG flow. Since its derivative approaches to zero in Fig 1(b), we see that the renormalized entropy converges into a certain value ($\approx - 435$) as $l \to \infty$.  }
\label{number}
\end{center}
\end{figure}

Now, let us determine to the coefficients, $a_0$, $a_1$, and $b_1$. Since the analytic evaluation of the integral in \eq{res:stripenten} is not allowed, we can only determine these values numerically. In Fig. 1, we depict the renormalized entropy and its slope relying the subsystem size where we take $d=3$, $L=1000$, $G=1$, and $z_h=1$ for simplicity. The leading correction in the IR limit is given by $a_0 \approx -435$ approximately, which is irrelevant to the RG flow. To determine what the next quantum correction is, we should first know whether $b_1$ is larger than $(d-2)$ or not. If $b_1 > (d-2)$, the next quantum correction comes from the remnant of the UV entanglement entropy, $a_{UV} l^{-(d-2)}$. To see that, let us define the following test function and numerically calculate it
\be
S_T \equiv l^{d-1} \ \fr{d \bar{S}}{d l} .
\ee
If this value diverges in the IR limit, it indicates $b_1 < (d-2)$. Otherwise, $b_1 \ge (d-2)$. Furthermore, if the IR value of $S_T$ approaches to $- (d-2) \ a_{UV}$, it means $b_1>(d-2)$ because for $b_1 = (d-2)$ it should converge into another value, $- (d-2) (a_1 + a_{UV})$. The numerical result in Fig. 2 indicates $b_1 > (d-2)$. Therefore, the first quantum correction comes from the short distance quantum correlation near the entangling surface. The similar behavior also occurs for $d=4$ case. These results imply that the IR entanglement entropy approaches the thermal entropy, as mentioned before, and the quantum correction is rapidly suppressed by the $l^{-(d-2)}$ power.

\subsection{Ball-shaped entangling region}

In the previous section, we have shown that the entanglement entropy stored in a strip-shaped region reduces to the thermal entropy in the IR limit. In addition, we also found that, when the subsystem size increases, the quantum correction caused by the short distance quantum correlation is suppressed by $l^{-(d-2)}$ for a $d$-dimensional CFT. For a higher dimensional field theory theory, one can consider a different shape of the entangling surface, like a spherical one which can give rise to additional information associated with the free energy and central charge of a dual field theory. In this section, we will investigate the IR entanglement entropy accumulated in a ball-shaped region and its universality discussed in the previous sections. For describing a spherical entangling surface with a rotational symmetry, it is more convenient to parameterize the AdS black hole metric in terms of the spherical coordinate
\be
ds^2 = \fr{1}{z^2} \ls - f(z) dt^2 + d \r^2 + \r^2 d\O_{d-2}^2 + \fr{1}{f(z)} dz^2 \rs ,
\ee
where $\O_{d-2}$ indicates the solid angle of a $(d-2)$-dimensional unit sphere. Denoting the radius of the entangling surface as $l$, the range of $\r$ is limited to $0 \le \r \le l$. In this case, the entanglement entropy is given by
\be      \la{act:sphericalHEE}
S_E = \fr{\O_{d-2}}{4 G}  \int_\e^{z_0} d z \ \fr{\r^{d-2}  \sqrt{1 +  f \r'^2}  }{z^{d-1} \sqrt{f}}  ,
\ee
where the prime means a derivative with respect to $z$.

\begin{figure}
\begin{center}
\vspace{0cm}
\hspace{-0.5cm}
\subfigure[]{ \includegraphics[angle=0,width=0.4\textwidth]{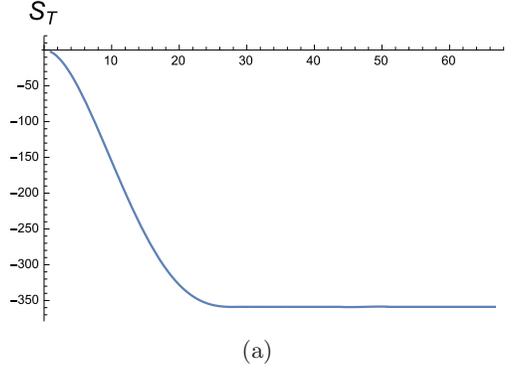}}
\vspace{-0cm}
\caption{\small The value of the test function $S_T$. In the IR limit, it approaches to $S_T \approx -359$ which is almost equal to the value of $- (d-2) \ a_{UV}$.}
\label{number}
\end{center}
\end{figure}

In order to clarify the entanglement entropy in the IR region, let us think of boundary conditions satisfied by the minimal surface. First, $\r(z)$ at $z=0$ must approach $l$ because the entangling surface is located at the boundary denoted by $z=0$. Due to the rotational symmetry of the minimal surface, $\r(z)$ should vanish at the turning point, $\r(z_0)=0$. In addition, the smoothness of the minimal surface requires $\pa_z \r |_{z=z_0}= - \infty$ at the turning point. These constraints fix the leading behavior of $\r(z)$ near the turning point to be
\be      \la{ansatz:turningptsol}
\r(z) \approx \ls z_0 - z \rs^\n \ c(z)  + \cdots ,
\ee
with $0< \n <1$, where $c(z)$ must be regular at the turning point. Near the turning point, because $\r' \to -\infty$, the above integral reduces to
\be   \la{ans:behaviorattp}
 \fr{\O_{d-2}}{4 G}  \int  d z \ \fr{\r^{d-2}  \sqrt{\r'^2}  }{z_0^{d-1} }  \to
 \fr{\O_1}{4 G z_0^{d-1}}  \int  d \r \  \r^{d-2}   .
\ee
From this fact, we can rewrite the excited state entanglement entropy as the following form
\be        \la{act:fsphericalHEE}
S_E = \fr{\O_{d-2}}{4 G z_0^{d-1}}  \int_0^{l} d \r \ \r^{d-2}
+\fr{\O_{d-2}}{4 G }  \int_\e^{z_0} d z \   \fr{\r^{d-2}  \ls z_0^{d-1} \sqrt{1 +  f \r'^2} -  z^{d-1}  \sqrt{ f \r'^2} \rs }{z_0^{d-1}  z^{d-1} \sqrt{f} }   .
\ee
In the IR limit ($l \to \infty$ and $z_0 \to z_h$), this decomposition shows that the first term reduces to the thermal entropy corresponding to the Bekenstein-Hawking entropy of the dual black hole. On the other hand, the remaining quantum correction is finite up to a UV divergence similar to the previous strip case. This finiteness of the quantum correction again implies that the thermal entropy naturally appears as the leading contribution to the IR entanglement entropy regardless of the dimension and shape of the entangling surface.

In the UV limit, the entanglement entropy stored in the ball-shaped region shows a totally different behavior depending on the dimension, so that we should be careful to investigate the entanglement entropy in the UV region. However, the dimension of the entangling surface is not crucial when investigating the IR behavior. In this work, thus, we focus on the cases with $d=3$. If we set $f=1$, the resulting geometry becomes a pure AdS space dual to a $(2+1)$-dimensional CFT. The entanglement entropy evaluated in this background geometry corresponds to the ground state entanglement entropy of the dual CFT, which is determined in terms of the subsystem size
\be
S_g = \fr{\O_1}{4 G}  \fr{l}{\e}- \fr{\O_1}{4 G} .
\ee
The entanglement entropy for the excited state can be determined by the black hole geometry with $f = 1 - z^3/z_h^3$. From \eq{act:sphericalHEE}, the minimal surface configuration is governed by
\be     \la{eq:sphericalEEeq}
0 =   \rho'' +\frac{2 \left(z_h^3-z^3\right)  \rho '^3}{ z_h^3 \ z}
   -\frac{\rho '^2}{\rho }
+ \frac{\left(4 z_h^3-z^3\right) \rho '}{2 z \left(z_h^3-z^3\right)}
-\frac{z_h^3}{\rho  \left(z_h^3-z^3\right)} .
\ee
Substituting the expected ansatz in \eq{ansatz:turningptsol} into this equation of motion, the solution has the following perturbative expansion
\be      \la{eq:nearz0}
0 =\frac{2 \n ^3  c_0^3 }{z_0} \left(z_0-z\right){}^{3 \n -3} -c_0 \n  \left(z_0-z\right){}^{\n
   -2}-\frac{1}{c_0} \left(z_0-z\right){}^{-\n } + \cdots ,
\ee
where $c_0$ indicates the value of $c(z)$ at the turning point and the ellipsis involves less divergent terms. In order to determine $\n$, let us consider the following three cases:

\begin{itemize}
\item For $\n > 1/2$, the last two terms in \eq{eq:nearz0} are more dominant. Cancelling of these two terms in \eq{eq:nearz0} fixes $\n$ to be $1$.  However, since $0< \n <1$ from the smoothness of the minimal surface, there is no $\n$ satisfying the equation of motion and smoothness of the minimal surface simultaneously.

\item For $\n < 1/2$, the first and third term are dominant. This means $\n=3/4$ which is inconsistent with the assumption, $\n < 1/2$. Thus, there is also no solution.

\item For $\n = 1/2$, the first two terms are dominant. In order to satisfy \eq{eq:nearz0} at leading order, $c_0$ must be
\be        \la{res:valueofc0}
c_0 = \fr{ \sqrt{2 \ z_0 }}{\sqrt{1 - z_0^3/ z_h^3}}  .
\ee

\end{itemize}

Near the turning point parameterized by $z_0 - \d \le z \le z_0$, the corresponding distance in the $x$-direction is given by $0 \le x \le l_x$ with
\be     \la{res:sizedelta}
l_x \sim \fr{ \sqrt{2 \ z_0 \ \d }  }{\sqrt{1 - z_0^3/ z_h^3}}  .
\ee
Substituting the solution \eq{ansatz:turningptsol} with the above $c_0$ into \eq{act:sphericalHEE}, we can find that the renormalized entropy near the turning point behaves as
\be
S_E = \fr{\O \d}{12 G ( z_h -z_0 )} + \cdots .
\ee
Rewriting it by using  \eq{res:sizedelta}, we finally reaches to
\be			\la{res:ballthermalen}
S_E = \fr{ \O_1  l_x^2}{8 G z_h^2} + \cdots ,
\ee
which is the main contribution to the IR entanglement entropy. In the IR limit, since $l_x \approx l$, this result exactly reproduces the thermal entropy corresponding to the Bekenstein-Hawking entropy of the dual black hole. This result indicates that the IR excited state entanglement entropy is thermalized from the center of the entangling region. This result becomes more manifest when we rewrite the renormalized entropy as the form in \eq{act:fsphericalHEE}
\be     \la{res:sphericalHEE3}
\bar{S} = \fr{\O_{1}}{4 G z_0^{2}}  \int_0^{l} d \r \  \r   +
\fr{\O_{1}}{4 G}  \ls \int_\e^{z_0} d z \ \fr{\r  \ls   z_0^{2} \sqrt{1 +  f \r'^2}    -    z^{2}  \sqrt{f \r'^2}   \rs}{z_0^{2}  z^{2} \sqrt{f}}
-  \fr{l}{\e} \rs + \fr{\O_1}{4 G}.
\ee
In the IR limit the first term corresponds to the thermal entropy appearing in \eq{res:ballthermalen}, while the remaining terms represent the quantum correction which is finite in the entire region of $l$. Since the thermal entropy is dominant in the IR limit, the IR entanglement entropy of the ball-shaped region reduces to the thermal entropy, as mentioned before.

\section{Universal thermal entropy from the IR entanglement entropy}

In the previous sections, we showed that the main contribution to the IR entanglement entropy comes  from the thermal entropy regardless of the shape of the entangling surface for a $d$-dimensional CFT. In order to figure out this feature holographically, an important ingredient is the existence of the  horizon in the dual geometry. The minimal surface extended near the horizon, corresponding to the center of the entangling region, leads to the most of the entanglement entropy in the IR limit. The existence of the horizon is a natural property of a black hole solution even for non-AdS geometries. Applying the gauge/gravity duality, therefore, one can expect that the thermal entropy universally appears in the IR entanglement entropy even for non-conformal field theories. In order to check the universal feature of the IR entanglement entropy, in this section we will show holographically that the thermal entropy leads to the main contribution to the IR entanglement entropy even for non-conformal relativistic field theories.

For a $(d+1)$-dimensional gravity theory, an almost general black hole metric can be represented as
\be            \la{met:generalmetric}
ds^2 =  \fr{1}{z^2} \ls - e^{2 A(z)} f(z) dt^2 + e^{2 B(z)} \d_{ij} dx^i dx^j + \fr{e^{2 C(z)}}{f(z)}  dz^2 \rs ,
\ee
where $i=1, \cdots, d-1$ and $f(z)$ indicates the black hole factor. Depending on the detail of the gravity theory, the black hole factor can have several roots. We denotes the largest root as $z_h$ which is called the black hole horizon. For convenience, the black hole factor can be further rewritten as the following form
\be
f(z) = \ls 1 - \fr{z}{z_h} \rs F(z) ,
\ee
where $F(z)$ must be regular for $0 \le z \le z_h$ and approaches to $1$ as $z \to 0$. The other unknown functions, $e^{2 A(z)}$, $e^{2 B(z)}$ and $e^{2 C(z)}$, are also regular except $z=0$. Using these facts, the Bekenstein-Hawking entropy reads from the area law
\be         \la{res:themalentropy}								
S_{th} = \fr{V_{d-1}}{4 G} \fr{e^{(d-1) B(z_h) } }{z_h^{d-1}} ,
\ee
where $V_{d-1}$ indicates a regularized volume in ${\bf R}^{d-1}$. Following the gauge/gravity duality, the Bekenstein-Hawking entropy can be reinterpreted as the thermal entropy of the dual QFT. In this case, the area of the black hole proportional to $V_{d-1}$ can be mapped to the volume of the dual QFT. This fact is important to identify the Bekenstein-Hawking entropy with the thermal entropy because the thermal entropy of a usual thermal system should be an extensive quantity. Above we assumed a rotational invariance in ${\bf R}^{d-1}$. We can further generalized it to a more general black hole solution breaking such a rotational symmetry. However, since breaking of the rotational invariance does not affect our study on the universality of the IR entanglement entropy, we concentrate on the above black hole metric.

Note that we can set $e^{2 C(z)}=1$ without loss of generality because of the diffeomorphsim invariance. In this case, the resulting metric and its dual field theory can be classified by $A(z)$ and $B(z)$ as follows:

\begin{itemize}
\item{For $e^{2 A(z)} = e^{2 B(z)} =1$, the metric reduces to that of the AdS black hole studied in the previous sections. The dual field theory is conformal.}

\item{For $e^{2 A(z)} \ne e^{2 B(z)} =1$, it reduces to the Lifshitz black hole which breaks the boost symmetry in the $t-x^i$ plane. The resulting dual field theory is a non-relativistic field theory with a scale invariance \cite{Taylor:2008tg,Park:2013goa,Park:2013dqa}.}

\item{For $e^{2 A(z)} = e^{2 B(z)} \ne 1$, it leads to a black hole on the hyperscaling violation geometry which has no scale symmetry. The dual field theory can be identified with a relativistic quantum field theory without a scale symmetry \cite{Goldstein:2009cv,Charmousis:2010zz,Kulkarni:2012re,Park:2013ana,Kulkarni:2012in}.}

\item{For $e^{2 A(z)} \ne 1$, $e^{2 B(z)} \ne 1$ and $e^{2 A(z)} \ne e^{2 B(z)} $, it is the combination of the previous two cases. In this case, the scale and boost symmetry are broken and the dual field theory is given by a non-relativistic theory without a scale symmetry.}

\end{itemize}

For a strip-shaped region,  the entanglement entropy is governed by
\be
S_E = \fr{L^{d-2}}{4 G} \int_{-l/2}^{l/2} dx \ \fr{e^{(d-2) B}   \sqrt{f e^{2 B}  + z'^2}  }{z^{d-1} \sqrt{f}} .
\ee
Using the conserved quantity caused by the translational symmetry in the $x$-direction, the width of the strip and the entanglement entropy are parameterized as functions of the turning point, $z_0$,
\be
l &=& 2 \int_0^{z_0} dz \ \fr{z^{d-1} \ e^{(d-1) B_0}  }{e^B \ \sqrt{f} \ \sqrt{e^{2 (d-1) B } z_0^{2(d-1)} - e^{2 (d-1) B_0 } z^{2(d-1)}}}   ,   \la{res:integrall}   \\
S_E &=&   \fr{L^{d-2}}{2 G} \int_0^{z_0} dz \ \fr{z_0^{d-1} \ e^{(2d-3) B_0} }{z^{d-1} \ \sqrt{f} \ \sqrt{e^{2 (d-1) B } z_0^{2(d-1)} - e^{2 (d-1) B_0 } z^{2(d-1)}}}  \la{res:gHEEstrip} ,
\ee
where $B_0$ implies the value of $B(z)$ at $z=z_0$. Here the range of the turning point is restricted to $0 \le z_0 \le z_h$  and $1/z$ corresponds to the energy scale of the dual QFT. This relations imply that $z_0 =0$ and $z_0 =z_h$ can map to a UV and IR limit of the dual QFT. When $z_0$ approaches to $0$, the integral in \eq{res:integrall} automatically vanishes. On the other hand, if $z_0$ approaches $z_h$ the integrand of \eq{res:integrall} gives rise to a simple pole. Performing the integral in\eq{res:integrall} near $z_0 = z_h$ yields the following relation at leading order
\be
l \approx z_0 \log \ls z_h - z_0 \rs .
\ee
This implies that the width of the strip diverges logarithmically in the IR limit. Rewriting the entanglement entropy by using \eq{res:integrall}, we can find the following form
\be      \la{res:genstripHEE}
S_E = \fr{l L^{d - 2}}{4 G} \fr{e^{(d - 1) B_0}  }{z_0^{d - 1}} + \fr{L^{d-2}}{2 G z_0^{d-1} } \int_\e^{z_0} dz \
\fr{ \sqrt{e^{2 (d-1) B } z_0^{2(d-1)} - e^{2 (d-1) B_0 } z^{2(d-1)}} }{z^{d-1}  e^{B}   \sqrt{f} } .
\ee
 Noting that the volume of the strip is given by $V_{d-1} = l L^{d-2}$, we can easily see that in the IR limit ($l \to \infty$), the first term exactly reduces to the thermal entropy of the dual field theory. Ignoring the UV divergence which is absent for the renormalized entropy, the quantum correction part gives rise to the regular contribution. As a consequence, since the first term is dominant in the IR region, the IR entanglement entropy reduces to the thermal entropy, as expected before.

Now, let us further study the entanglement entropy accumulated in a ball-shaped region. Due to the rotational symmetry of the ball-shaped region, it is more convenient to rewrite the metric in \eq{met:generalmetric} as the following form, which makes the rotational symmetry manifest
\be
ds^2 =  \fr{1}{z^2} \ls - e^{2 A(z)} f(z) dt^2 + e^{2 B(z)}  d \r^2 + e^{2 B(z)}   \r^2 d\O_{d-2}^2+ \fr{1}{f(z)}  dz^2 \rs .
\ee
On this background metric, the entanglement entropy reads
\be
S_E = \fr{\O_{d-2}}{4 G} \int_0^l d\r \ \fr{e^{(d-2) B} \r^{d-2} \sqrt{e^{2 B} f + z'^2} }{z^{d-1} \sqrt{f}} .
\ee
For a pure AdS geometry with $B = 0$ and $f=1$, the exact configuration of the minimal surface has been  known as $z = \sqrt{l^2 - \r^2}$. However, if $B \ne 0$ or $f \ne 1$, it is not easy to find an exact solution. In spite of this fact, there are several constraints the solution must satisfy. First, the entangling surface is located at the boundary, so that the solution must have $z(l) = 0$. Another constraint is that $z$ has a turning point at $\r=0$ due to the rotational symmetry. Furthermore, the smoothness of the minimal surface requires to be $z'=0$ at the turning point. Due to these constraints, the entanglement entropy near the turning point should be approximately proportional to
\be
\fr{\O_{d-2} e^{(d-1) B_0}}{4 G z_0^{d-1}} \int_{z \approx z_0} d\r \ \r^{d-2} .
\ee
This behavior becomes manifest when we rewrite the above entanglement entropy as the following form
\be      \la{res:ballentangle}
S_E &=& \fr{\O_{d-2} e^{(d-1) B_0}}{4 G z_0^{d-1}} \int_0^l d\r \ \r^{d-2} \nn
&& +  \fr{\O_{d-2} }{4 G z_0^{d-1}} \int_0^l d\r \ \fr{ \r^{d-2}  \ls z_0^{d-1} e^{(d-1) B} \sqrt{f + e^{-2 B}  z'^2} - z^{d-1} e^{(d-1) B_0 } \sqrt{f} \rs }{z^{d-1} \sqrt{f}} .
\ee
Noting that the volume of the ball-shaped region is given by $V_{d-2}= \O_{d-2} \int_0^l d\r \ \r^{d-2}$, one can see that in the $z_0 \to z_h$ limit the first integral is exactly reduced to the thermal entropy which diverges as $l \to \infty$. Ignoring the UV divergence, the second term corresponding to the quantum correction is always  finite. Similar to the strip case, the IR entanglement entropy of the ball-shaped region exactly reduces to the thermal entropy in the IR limit.

Intriguingly, all results studied in this work show that the IR entanglement entropy reduces to the thermal entropy in the IR limit regardless of the microscopic detail. This implies that, through the generalized temperature defined in this work, the macroscopic thermodynamic law can be derived from the thermodynamics-like law of the quantum entanglement entropy in the IR limit.


\section{Discussion}

In the quantum information theory, it has been shown that quantum information evolves into the thermal entropy via a unitary time evolution \cite{Popescu:2006,Sagawa:2016}. This fact implies that there exists a connection between the quantum entanglement entropy and the thermal entropy. Thus, clarifying such a connection plays a crucial role for understanding the microscopic origin of various macroscopic and thermodynamic phenomena. In this work, we introduced the generalized temperature, which is valid even in the IR region and required to describe the RG flow correctly, and then investigated holographically how the quantum entanglement entropy evolves into the thermal entropy along the RG flow.

In the UV regime, the entanglement entropy has nothing to do with the thermal entropy. This becomes manifest from the UV behavior of the generalized temperature. In the UV region, the leading contribution to the generalized temperature is inversely proportional to the subsystem size, while the thermodynamic temperature must be independent of the system size. Due to this fact, although the similar thermodynamics-like relation governs the UV entanglement entropy, it cannot be reinterpreted as the thermodynamic law of a real thermal system. However, this UV story dramatically changes in the IR regime.

In the IR limit, the entanglement entropy can be decomposed into two parts. One is the dominant contribution caused by the thermalization of the excited state entanglement entropy, which leads to the thermal entropy corresponding to the Bekenstein-Hawking entropy of the dual black hole geometry. The other is the remaining quantum entanglement near the entangling surface which is always smaller than the thermal entropy in the IR region. In addition, the generalized temperature approaches the thermodynamic temperature corresponding to the Hawking temperature of the dual black hole geometry. These IR features of the entanglement entropy and the generalized temperature universally occur regardless of the microscopic detail and the shape of the entangling surface. These facts imply that the thermodynamics-like law governed by the entanglement entropy evolves to the real thermodynamic law governed by the thermal entropy. The universal IR feature has been checked in various holographic models, so that it would be interesting to derive the same IR universality on the quantum field theory side, for example, a variety of the low-dimensional Ising models \cite{Kim:2016ayz,Kim:2016hig}.

\vspace{0.5cm}

{\bf Acknowledgement}

C. Park was supported by Basic Science Research Program through the National Research Foundation of Korea funded by the Ministry of Education (NRF-2016R1D1A1B03932371) and also by the Korea Ministry of Education, Science and Technology, Gyeongsangbuk-Do and Pohang City. K.-S. Kim was supported by the Ministry of Education, Science, and Technology (No. NRF-2015R1C1A1A01051629 and No. 2011-0030046) of the National Research Foundation of Korea (NRF) and by TJ Park Science Fellowship of the POSCO TJ Park Foundation. This work was also supported by the POSTECH Basic Science Research Institute Grant (2016). We would like to appreciate fruitful discussions in the APCTP Focus program ``Lecture Series on Beyond Landau Fermi Liquid and BCS Superconductivity near Quantum Criticality" (2016).



\end{document}